# Living in the tensions: Investigations of gender performativity in STEM

Smith Strain [1], Noah Leibnitz [2], Reagan Ruben [1], Yangqiuting Li [2] and Eric Burkholder [1]

[1]*Department of Physics, Auburn University, Auburn, AL 36849*
[2]*Department of Physics, Oregon State University, Corvallis, Oregon 97331*

In this work, we present the results of semi-structured interviews with four women to explore how they perceive themselves with respect to three gender constructs (femininity, masculinity, androgyny), and how they believe others perceive them. All the women highlighted the performative nature of gender in science, technology, engineering, and mathematics (STEM), citing (1) stereotypes that women are not analytical thinkers, or femininity being associated with "being stupid"; (2) the pressure to conform to the masculine norms of STEM, and (3) a pressure to perform to prove that they belong in STEM. Some of these women aligned their own perceptions of their gender with these norms, while others expressed frustration with the tension between their gender and how that is perceived by peers in STEM. This work suggests that conceptualizing gender as performance is a useful lens for understanding the oppression and underrepresentation of women and gender minorities in STEM.

Note: Smith Strain and Noah Leibnitz are co-first authors.

## I. INTRODUCTION

Physics continues to be a male-dominated field, having one of the lowest representations of women of all science, technology, engineering, and mathematics (STEM) disciplines [1,2]. Although women earned about 65% of all bachelor's degrees in the US in 2020, they accounted for only 25% of undergraduate degrees in physics [1,2]. This gender disparity extends to higher academic levels, with women earning 21% of physics doctoral degrees, holding 18% of postdoctoral positions, and representing 14% of faculty members [2,3]. Numerous studies in physics education research have addressed these disparities, seeking to promote inclusivity, equity, and diversity [4-15]. They have examined various factors contributing to gender disparities in areas such as academic performance and motivation [6,10,14,16-23], including stereotype threat [12,24], the scarcity of role models [4,25,26], and systemic biases [27,28]. Efforts to bridge these gaps have included psychological interventions [19,29], active learning strategies [16,30], and revised assessment methods [31,32] to foster more inclusive educational environments.

Most previous studies have used binary gender classifications, which may have limited the breadth of research questions and the depth of findings [33-37]. Not only do these classifications erase the experiences of nonbinary individuals, but they obscure the nuances afforded by considering gender as a *spectrum* of identities and behaviors. Recent research [38,39] using gradational measures of femininity, masculinity, and androgyny has shown significant variation within binary gender categories, offering new insights into gender dynamics in physics. For example, one prior study revealed that self-identified women in introductory physics courses, especially in courses where they were underrepresented, tended to view themselves as more masculine and androgynous–and less feminine–than how they thought others perceived them (known as "reflected appraisals" of gender) [39]. The study showed that these discrepancies were closely related to students' gender stigma consciousness, which refers to individuals' awareness and recognition of the stigmatization and negative stereotypes associated with gender [39].

To supplement these prior findings, we conducted semi-structured interviews with students in introductory physics courses to qualitatively investigate how they perceive their own femininity, masculinity, and androgyny, and how they believe they are perceived by others. Our two research questions are **1. How do self-identified women's perceptions of STEM culture relate to their self-perceived and reflected appraisals of femininity, masculinity, and androgyny? 2. To what extent—and how—do STEM environments cause tensions between gender identity and expression for women?** Our interview data provide insights into students' gender performance, the impact of the STEM learning environment on their identities and expressions, and their reasoning behind discrepancies between their own identities and expectations from others.

## II. THEORETICAL FRAMEWORK

We adopted Judith Butler's theory of gender performativity [40] as a framework for this study. According to Butler, gender is a social and cultural construct that is created and maintained through "performances" [40] such as a series of actions, gestures, and speech. These performances are not entirely individual choices or expressions of an innate gender, but instead are shaped by societal norms, expectations, and discourses [40]. In addition to the performative aspects of gender, prior studies have also emphasized the interactive nature of gender [41-43]. Individuals "do" their gender by enacting patterns of behavior that are socially understood to be feminine or masculine (or neither), and their gender is simultaneously "determined" by others who interpret those enactments [41]. Research has revealed that while there is a connection between how individuals perform their gender and how others perceive it, discrepancies can arise between individuals' self-perception of their gender and how others interpret it [42], which can cause cognitive and emotional strain [43]. Previous findings have also suggested that gender stigma consciousness has a mediating effect on misalignment between women's self-perceived and reflected appraisals of their femininity/masculinity/androgyny in physics courses [39]. Considering this, we set out in this study to explore how the STEM environment interacts with women's experience and performance of their gender with the goal of discerning how STEM culture might be influencing such misalignment. We constructed narratives outlining the women's viewpoints and behaviors and connected these stories back to the theory of gender performativity in interpreting the data.

## III. METHODOLOGY

This research was determined exempt from review under (University's name is redacted for review) IRB protocol 23-349. Interview participants were chosen based on their responses to a survey taken in the first physics recitation section of the Fall 2023 semester at a large research university. This survey was administered to students in all introductory physics courses including calculus-based, and algebra-based physics 1 and 2. We identified students who (1) consented to participate in the study, (2) self-identified as something other than a white, cisgendered man and (3) indicated discrepancies between their self-perceived and reflected appraisals of their femininity/masculinity/androgyny. Data for (3) were collected using student responses to the following questions [38], which were Likert-scale questions ranging from 0 (Not at all) to 6 (Very) as shown in Fig. 1 and Fig. 2.

In general, how do you see yourself? (Please answer all three scales).

|  | Not at all | 1 | 2 | 3 | 4 | 5 | Very |
|---|---|---|---|---|---|---|---|
| Feminine | ○ | ○ | ○ | ○ | ○ | ○ | ○ |
| Masculine | ○ | ○ | ○ | ○ | ○ | ○ | ○ |
| Androgynous | ○ | ○ | ○ | ○ | ○ | ○ | ○ |

**FIG. 1.** Survey questions for self-perceived femininity, masculinity, and androgyny.

In general, how do most people see you? (Please answer all three scales).

|  | Not at all | 1 | 2 | 3 | 4 | 5 | Very |
|---|---|---|---|---|---|---|---|
| Feminine | ○ | ○ | ○ | ○ | ○ | ○ | ○ |
| Masculine | ○ | ○ | ○ | ○ | ○ | ○ | ○ |
| Androgynous | ○ | ○ | ○ | ○ | ○ | ○ | ○ |

**FIG. 2.** Survey questions for reflected appraisal of femininity, masculinity, and androgyny.

These students were invited to participate in an interview about their survey responses during the semester they were taking their physics course. One interviewer was a white cisgendered woman in engineering, and the other was an Asian cisgendered woman in physics. Participants were given the option to speak with whichever interviewer they felt most comfortable with and received a $25 gift card for participating. In the interview, the participants were asked to answer the survey questions again. They were also asked to explain their choices, particularly when there were discrepancies between reflected appraisals and self-perceptions, or when their choices differed from those they gave when the survey was completed in class. They were asked about other potential influences on these answers such as the classroom environment, their interactions with peers, and the university environment more broadly. They were also asked about steps the institution might take to improve inclusion and consideration of more diverse expressions of gender.

As this research was exploratory, we adopted a phenomenological approach [44] to analyze the data. N. L. and R. S. S. initially listened to each interview and took notes on interesting themes that they noticed, paying particular attention to responses that were consistent across the different cases, responses that were unique and focused on intersections with other aspects of identity, or responses that contradicted the responses of other participants. The research team then met as a group to discuss particularly interesting cases. There were ten total participants; we selected four of them for deeper analysis as they provided answers that were more focused on how STEM impacted their answers. We constructed narratives for each of these cases in line with Standpoint Theory—which emphasizes the importance of marginalized groups, whose social positioning has provided with unique perspectives—to center the lived experiences of these participants [45,46]. The research team then met to discuss preliminary themes, which were resolved by conversations between N. L. and R. S. S.

The authors of this study identify as cis-gendered men and women, members of the queer community, non-White domestic scholars, and international scholars. Three of the authors were at the research site during the time of the study, one author had left this institution, and another author was at a different institution. Two of the authors were raised in the same geographic region as the study site, while the others come from other regions of the world. This provided a complex blend of perspectives that allowed for multiple different framings and interpretations of the data. For example, two of the researchers (the interviewers) were women at the research site and related directly to the experiences of the students they interviewed. Other members of the research shared partially overlapping identities with the participants that allowed them to draw from their experiences as women of color or queer people. In the analysis, we were careful to consider all perspectives while focusing more intently on the perspectives of those with aspects of minoritized identity that overlapped with the study participants.

## IV. RESULTS

The results are presented in narrative format to provide more fidelity to the lived experiences of the participants. The four participants were Aarya, Mia, Ana, and Hannah (all self-identified cisgendered women); these are pseudonyms which we chose to reflect both the ethnic and gender identities of the participants. Overall, these women's responses are consistent with previous findings that there are discrepancies between students' self-perceived and reflected appraisals of their femininity/masculinity/androgyny [39]. The interviewed women's self-perceptions of their gender are typically shaped by socialized notions of gender and pervasive Western gender stereotypes. Many of them reported tensions between their own identities and the expectations from others, due to the male-dominated and directed STEM environment.

### 1. Aarya

Aarya felt that she could only perceive herself as a 4/6 on the femininity scale due to social norms of femininity. *"It's kind of like there's like one path you can take… it's kind of how femininity is defined right now"* she says, citing a traditional wife ("trad-wife") lifestyle that is associated with stereotypical femininity. As a computer engineer, she often conveys, consciously or not, that she believes she is perceived as more masculine than she actually feels (she only perceived herself as a 2/6 on the masculinity scale). Her self-perceived androgyny was a 3/6, though she characterized this in terms of performance and not an internalized feeling.

When asked about her reflected appraisal of femininity (which she rated as a 5/6), she explained that *"I come off as feminine because, just like the way I dress, the way I act, so like I'm the only one that knows how I actually feel about my gender,"* indicating a tension between how she feels about her own femininity and how people perceive her femininity. Aarya's reflected appraisal of masculinity (2/6) offered more insight into how she perceives gendered stereotypes of computer engineers more broadly. *"Well, this one I kind of thought about like, it was more like how I thought about my major, kind of like, I don't think I don't see a lot of women in computer engineering, like more comp sci.-based classes."* Because of this, she believes *"others would perceive [her] as more masculine, if [she took up] …positions that are normally…more male-oriented."* She felt that her self-perception and reflected appraisal of androgyny were the same, referencing the way that she dresses and looks, which was also her justification for her perceived femininity.

When asked if students in her major would answer these questions in the same way, Aarya explained that she would expect women in computer science or software classes to answer the same way that she did, but that environment also plays a key role. *"It depends on, like, who you surround yourself with, so like, if it were another major that is more female-oriented, they would obviously answer differently,"* in alignment with Butler's theory [40]. Aarya also mentioned how her upbringing never forced her to think deeply about her gender identity: *"I've been raised [in the deep South] my whole life, so it's just not a thing that I've like grown up—like gender identity is like not really talked about in school and stuff,"* indicating the potential for unique regional and cultural influences on perceptions of gender.

### 2. Mia

Mia's self-perceived and reflected appraisals of her masculinity (2/6) were equal, and she frequently explained her ratings using behaviors and traits traditionally associated with STEM. For example, when asked to explain her reported masculinity ratings, she noted that *"…obviously, like, I'm in STEM, which isn't masculine, but all the people in physics are guys. So like, it makes me feel a bit more on the perceived masculine thing,"* She also added that *"that's probably a perceived masculinity thing—like using math and stuff like that…"* indicating that just being in a STEM–especially math-intensive STEM–environment plays some role in her perception of her own gender.

In response to questions about the reasoning behind her ratings of her self-perceived and reflected appraisals of femininity (both 4/6), Mia drew attention to stereotypes and stigmas surrounding women in STEM. She felt that her peers in STEM often associated femininity with lower intelligence, resulting in her acting less feminine in STEM environments: *"in physics, I probably don't act as much feminine because people will think I'm acting stupid, even though that's not what femininity is."* This created a particular difficulty when trying to solicit help from male instructors, because *"sometimes with physics teachers, if I talk to them, like how I normally talk to people or how I talk to my friends, they'll think I'm stupid. Like, but if it's a girl she'll probably know what I'm talking about when I'm just like, trying to have a conversation—whereas guys will be like, 'what are you talking about?'"* Despite awareness of such stereotypes, Mia claimed not to *"really think about gender that much,"* which was a primary reason why she reported low levels of androgyny (both her self-perceived and reflected appraisal ratings were 2/6).

Mia further expressed feeling pressured to perform well academically because others perceived her—being one among few female physics majors—as a representative of women in STEM: *"[G]irls, like, as a whole, I think are more likely to be less confidently wrong. If we're not sure about a subject, we're not going to yell out the answer as much. And that makes it seem like we're not participating or not into their lectures…but even if we're confidently wrong, I still feel like it would feed the stereotype of us being, like, us being stupid. Yeah, yeah, we would just be like, it'd be wrong, because we were stupid. Not because we're trying, it would just be because we don't know what we're talking about."*

Furthermore, Mia expressed awareness of a pressure to succeed *beyond* what might be typically expected of the average student, saying that *"… some professors are really like, they really want women in STEM. So, they're going to act a little bit differently to female students, you know, like, they'll really, like try to help them out and like, they'll perceive them like possibilities and STEM, whereas people in my class probably just see me as another girl in a STEM class"* Mia perceived the culture in STEM—even well-intentioned behaviors aimed at supporting and elevating women—to create a pressure of expectation. In this way, a tension arose between her personal interests and motivations and those she felt were implicitly placed upon her because of her gender.

### 3. Ana

Ana's explanations of her answers to the survey questions were concise and rooted in a "feeling." For example, when asked why she chose 3/6 as her self-perceived femininity, Ana replied with *"I guess [I put] that because I really feel very neutral."* Despite this, she indicated that she thought others perceived her to be highly feminine (5/6). Intriguingly, Ana felt unable to make any further ratings regarding her masculinity or androgyny, citing a lack of ability to "identify with that side of the spectrum." However, Ana's responses to subsequent questions prompted a deeper discussion of her and her friends' experiences as women in STEM. When asked about the experiences of women in her major, she indicated that *"when I try to participate or anything in the class, [the male students] always put me aside, because as a female in the engineering side, they don't perceive a female as intelligent."* Ana alludes to feeling the need to adopt masculine traits and behaviors not only in the classroom, but in other areas of her life as a student. Ana said that *"more or less, I have to put on that, like, masculine type to be more included,"* and suggested this as a type of 'survival mechanism' (not a direct quote). Ana describes a time she was the only girl in her electrical lab and she had to join a group of men. She says that every time she would ask a question, she was written off as a "dumb girl". Describing how that experience made her feel, she explains that *"for [her], [it] was personally degrading in a way, because they said that it would have been easier if [she] were just a boy rather than a girl."* We asked her if she believed that other women would feel this need to put on a more masculine front, to which she responded *"yes… a lot of girls will do that. I have a friend who does that because she feels that girls are not included"*. Ana also expressed frustration: *"It's not fair for us, like we have to put on like a whole different side, a different character just to be excluded or to be included."* She seems to suggest that men do not have to alter their performance of gender to be viewed as successful.

### 4. Hannah

Hannah rated equally her self-perceived and reflected appraisals of femininity (6/10) and androgyny (4/10); we note that she used a 1-10 rating scale during the interview despite the questionnaire being a 0-6 scale. When asked to explain her survey responses, Hannah made a direct connection between involvement in STEM and lower femininity, commenting that *"being a STEM major, I'm not 100% feminine in the way I think about things."* She further acknowledged the existence of a relationship between her industrial engineering major and her gender expression when she commented: *"...being an engineering major and, like, I don't wear makeup regularly… I don't like, put a whole lot of effort into what I look like. I'm not always wearing the trendiest clothes; I'm not always painting my nails… So just, like, because of that, I would say [my femininity] is not, again, like to the max…"*

Hannah felt that possessing qualities often associated with STEM disciplines, such as being analytical, made her feel simultaneously less feminine and more masculine. She notably stated: *"I am feminine, but I just feel I also have qualities that are kind of not feminine. Like I'm very analytical, which I feel like isn't normally associated with females… And I feel like sadly, females are usually associated with…being more emotional, which I don't think is necessarily true. But I think because of that, like, just people noticing that I'm like, you know, smart, math oriented, that kind of thing. They think that, you know, oh, she's not the most feminine."*

Hannah's characterization of herself as someone with inclinations toward STEM also contributed to the difference she cited between her beliefs about her own masculinity and the way she believed others perceived her. She rated her self-perceived masculinity (1/10) higher than her reflected appraisal of her masculinity (0/10), matching a pattern observed previously [39]. She explained this, saying that *"just someone walking by me is not going to see that I'm like, really analytical, they're not going to see that I like, really love math and like, the stuff that some people consider to be more masculine."* This appears to indicate that Hannah accepts these stereotypes about STEM being more logical and analytical and thus associated with more masculine traits.

## V. DISCUSSION AND CONCLUSION

These narratives provide a rich illustration of the tensions that women navigate in STEM classrooms. In response to RQ1, we found a consistent theme of the women identifying with more stereotypically masculine traits–such as "analytical thinking," inclination toward mathematics, or being more career-focused due to the strong association between these traits and STEM. A few interviewees cited mere involvement in STEM as rationale for higher rates of masculinity and lower rates of femininity. There was also a strong theme of altering behaviors or appearance (consciously or not) to appear less feminine, such as not wearing make-up or altering speech patterns. Some of the women indicated that, despite conscious efforts, they still felt that they were looked down upon by their male peers or teachers. This reflects a "no-win" situation, where women simultaneously distanced themselves from femininity but were also still "too feminine" for STEM (RQ2).

In response to RQ2, we also found that many of these women report that acting more feminine altered the expectations placed upon them such as being unintelligent. Mia linked this to women being less likely to participate in class and suggested that, if a woman were to adopt masculine traits like being "confidently wrong," that would only reinforce the stereotype that women are less intelligent or competent in STEM. She also noted that well-intentioned faculty may contribute to this tension by trying to make the women be more involved or pushing them to succeed in the face of these unwelcoming classroom environments for the sake of increasing representation of women in STEM. Previous work has suggested that these kinds of gendered expectations may push women out of STEM [47].

These tensions around gender identity/expression appear to create a significant amount of excess cognitive load for these women, which is the hypothesized mechanism by which stereotype threat negatively impacts women's performance [48]. Though two of the women report not thinking about gender in terms of spectrums of femininity, masculinity and androgyny, this characterization of gender is at least partially reflective of their lived experiences. Indeed, being put into the STEM environment suddenly forced them to think about degrees of femininity and masculinity. Aarya suggested that this likely specific to traditionally male-dominated STEM majors and environments, which may explain why these trends are so prominent in physics classrooms. It may be the case that tensions between women's self-perceived and reflected appraisal of their gender both result from STEM culture and negatively impact their performance.

These narratives suggest that instructors need to be attuned not just to their own actions but monitoring their classroom for instances of gender discrimination–particularly on the part of male students. They may draw on some inclusive teaching strategies [49] such as elevating the voices of women and affirming their contributions to class discussions or putting women in more gender-diverse groups [50]. Suggestions from our interviewees include normalizing the use of preferred pronouns, allowing study/lab groups to form naturally, and being conscious of nonverbal communication. Another avenue that is showing promise is using *counternarratives* [51] to actively combat the notion of physics as the domain of white men, particularly given the nature of this stereotype in popular culture [52]. Instructors could also consider belonging interventions [53], though they should be aware of the limitations and nuances of these interventions in promoting cultural change in the classroom [54]. Future work should explore these tensions in non-physics environments and other institutional contexts to understand contextual factors that affect how women navigate STEM. It should also focus on students with non-binary gender identities, which may provide some insight as to whether it is femininity specifically that leads to these tensions for students, or just general feelings of "otherness" such as being androgynous or non-white. We plan to explore these issues further in forthcoming studies, as well as the long-term impacts of these tensions on student academic performance and motivational beliefs.